\documentstyle[12pt,epsf]{article}

\newcommand{\marke}[1]{\protect\label{#1} \fbox {\tt #1} \quad }
\renewcommand{\marke}{\label}

\title{The `diffusion' of light \\
and angular distribution in the laser \\
equipped with a multilobe mirror}
\author{Michael B.Mensky\\
{\small P.N.Lebedev Physics Institute, 53 Leninsky prosp., 117924 Moscow, Russia}\\
Alexander V.Yurkin\\
{\small General Physics Institute, 38 Vavilov Str., 117942 Moscow,
Russia}}
\date{}

\newcommand{\eq}[1]{(\ref{#1})}

\newcommand{\partderiv}[2]{\frac{\partial #1}{\partial #2}}

\newcommand{\al}{\alpha}

\newcommand{\be}{\begin{equation}}
\newcommand{\ee}{\end{equation}}
\newcommand{\ba}{\begin{eqnarray}}
\newcommand{\ea}{\end{eqnarray}}
\newcommand{\ban}{\begin{eqnarray*}}
\newcommand{\ean}{\end{eqnarray*}}

\newcommand{\bi}{\begin{itemize}}
\newcommand{\ei}{\end{itemize}}

\newcommand{\n}{{\bf n}}
\newcommand{\k}{{\bf k}}
\renewcommand{\a}{{\vec{\alpha}}}
\newcommand{\g}{{\vec{\gamma}}}
\newcommand{\r}{{\vec{\rho}}}
\newcommand{\e}{{\vec{\eta}}}
\newcommand{\ks}{{\vec{\xi}}}

\begin{document}
\maketitle

\begin{abstract}
The distribution of radiation is investigated for the modeless laser
having a multilobe mirror with the lobes (planes) inclined by small
angles to optical axis. It is shown that change of the direction
resulting from many passages of a ray through the optical system
including a multilobe mirror may be described as Brownian walk of a
point along the plane or equivalently as a solution of the
two-dimensional diffusion equation. Boundary conditions for the
diffusion equation may be approximately formulated as null conditions
at some angle which, if being reached during the walk, guarantees that
the ray escape from the optical system. In the framework of this
approximation an explicit formula for the distribution of the outgoing
ray in different angles is derived. After many passages through the
optical system the angular distribution tends to some universal
function. In the case of the round mirror it may be presented by the
Bessel function of order zero.
\end{abstract}


\section{Introduction}\marke{intro}

It is known \cite{NonSmooth} that the angular distribution of the
radiation of the laser with two plane-parallel mirrors is nonregular
(inhomogeneous) because of nonideal mirrors and inhomogeneity of the
active medium. This defect is absent in so called modeless lasers0
\cite{modeless} having no
longitudinal modes structure. Another configuration of modeless
laser was proposed in the work \cite{MultilobeMir}. In such a laser
one of the plane mirrors is replaced by an multilobe mirror having
the lobes (planes) slightly inclined to the optical axis
(Fig.\ref{FigOptics}~a). It was experimentally shown
\cite{MultilobeMir,Exper} that this replacement makes the angular
distribution of the laser radiation smooth (homogeneous).
There is no longitudinal modes in this laser too. Light is mixed
and scattered in the cavity of a multilobe mirror laser.

The form of the angular distribution in the multilobe laser was
analyzed in the work \cite{PrevTheory} by consideration of
geometry of the rays crossing the optical system with the
multilobe mirror many times. Such an analysis proves to be
complicated, may be carried out only for some simplest
configurations of the multilobe mirror and gives no
explicit formula for the angular distribution. In the present paper we
shall analyze the case of an arbitrary multilobe mirror and calculate
an approximate angular distribution in an explicit analytical form.
The idea of the method is following.

In each double passage through the optical system the ray is deflected by
a definite angle in one of a number directions depending on the
inclination of the mirror the ray was reflected by. Change of the
direction of the ray during multiple passage through the system have
therefore the character of random walk and may be described
mathematically by two-dimensional diffusion equation. Finiteness of
the aperture of the optical system may be approximately taken into
account by null boundary conditions at the boundary of some finite
region of the plane. The region is a disk if the optical system has
axial symmetry. In this case the problem may be easily solved by the
method of separation of variables so that and explicit expression in
the form of an infinite sum may be obtained for the resulting angular
distribution.

Moreover, since each term of the sum exponentially decays with the
number of passages increasing and the exponents are different, only one
of the terms (having the minimal exponent) dominates in the case of a
large number of passages. Thus, asymptotic (corresponding to large
number of passages) angular distribution of the radiation turns out to
be universal and is described by the Bessel function of order zero.

The reflection of the ray by the mirrors is considered in the
framework of geometric optics and therefore the method is applicable
only in the case of large Fresnel number, $a^2/\lambda L\gg 1$ (here
$\lambda$ is the maximal wave length in the wave packet, $a$ the
radius of the mirror and $L$ the distance between the multilobe mirror
and the opposite plane mirror closing the optical system). This means
that the ratio $\lambda/a$ should be much smaller than the ratio $a/L$.
We assume too that the thickness of the multilobe mirror $h\ll L$.

\section{Double passage of a ray through the
optical system}\marke{SectReflect}

If a ray of light which is directed along the unit vector ${\bf k}$
falls onto the plane mirror having the unit normal vector ${\bf n}$,
then the reflected ray is directed along the unit vector
(Fig.~\ref{FigOptics}~b)
\be
\k_1=\k-2(\k\n)\n
\marke{reflect}\ee
\begin{figure}[h]
\let\picnaturalsize=N
\def\picsize{1.5in}
\ifx\nopictures Y\else{\ifx\epsfloaded Y\else\input epsf \fi
\let\epsfloaded=Y
{\hspace*{\fill}
 \parbox{2in}{\ifx\picnaturalsize N\epsfxsize \picsize\fi
                                    \epsfbox{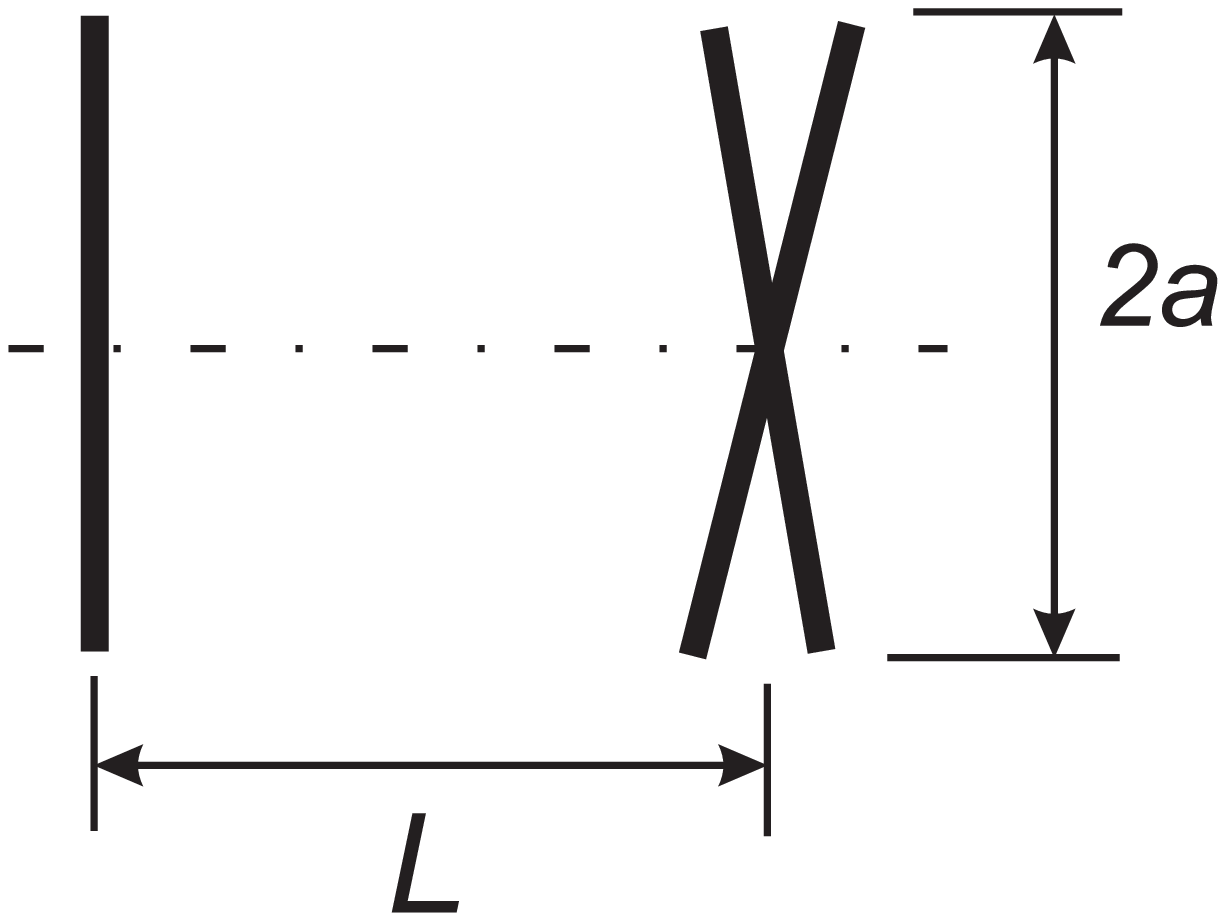}}\hfill
 \parbox{2in}{\ifx\picnaturalsize N\epsfxsize \picsize\fi
                                    \epsfbox{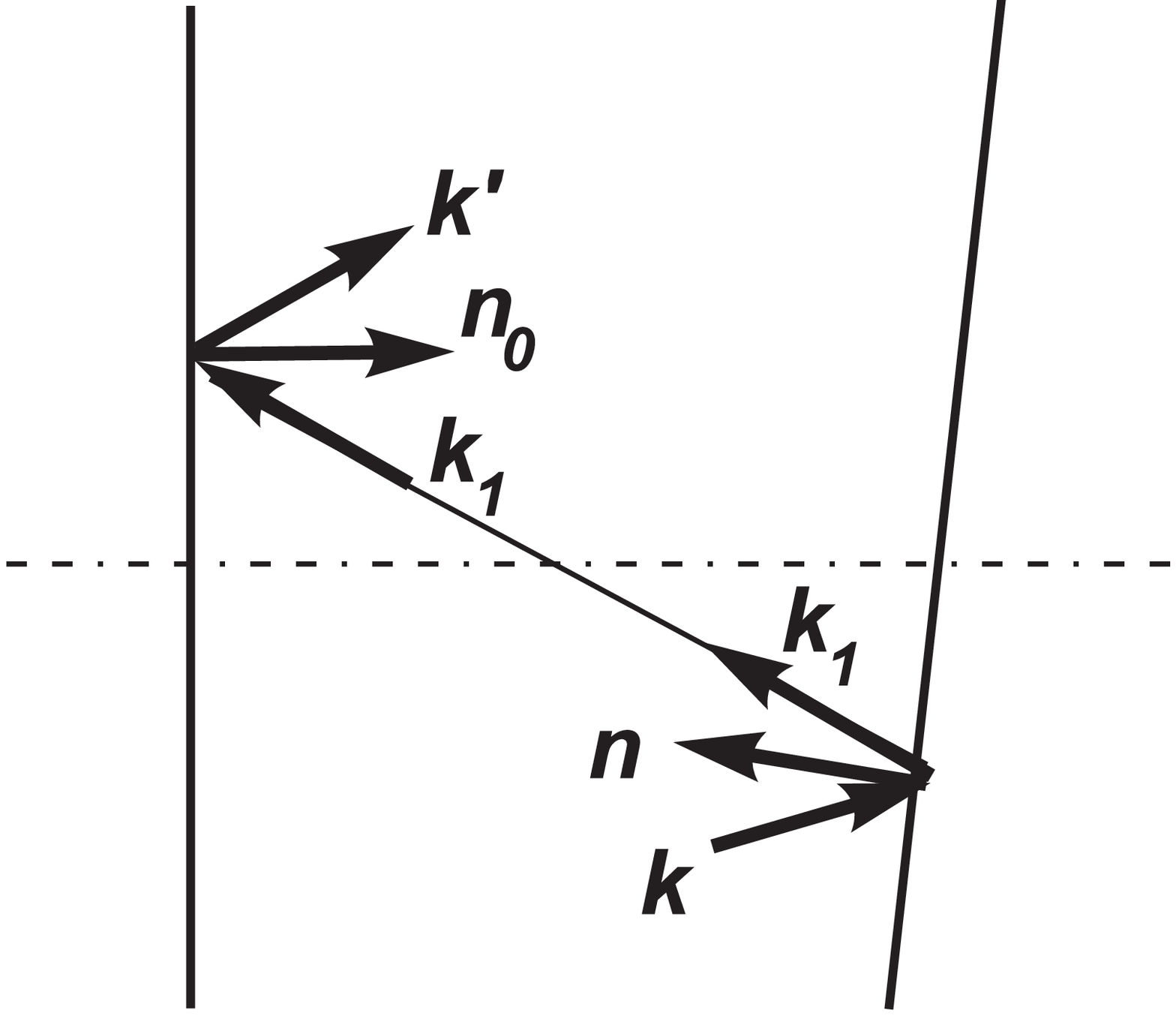}}\hspace*{\fill}}}\\
\par\vspace{0.5cm}
\hspace*{\fill}$a$\hfill$b$\hspace*{\fill}\\
\caption{An optical system equipped by a multilobe mirror: a)~The
general scheme; b)~the form of the ray reflected by a lobe of the
multilobe mirror and then by the closing mirror.}
\marke{FigOptics}
\end{figure}

Let $\n_0$ be the unit vector along the axis of the optical system.
Assume that the vector $\k$ characterizing the ray is close to $\n_0$
and the normal $\n$ to the mirror is close to $-\n_0$. Then we
can represent these vectors in the form
\be
\k = \n_0 + \a, \quad \n = -\n_0 + \g
\marke{smallAngl}\ee
where the vectors $\a$ and $\g$ are orthogonal to $\n_0$ and the terms
of the second order in $\a$ and $\g$ are omitted. The absolute values
of the vectors $\a$ and $\g$ are equal to the angles of inclination of
the ray and the mirror normal to the optical axis. The directions of
these vectors show in which direction the ray and the mirror are
inclined. Call therefore the vectors $\a$ and $\g$ `angle vectors' or
simply `angles'.

We shall assume that the inclination angles are small so that only
first order terms may be conserved in calculations. In this
approximation $\k\n=-1$ and therefore the reflection law \eq{reflect}
takes the form
\be
\k_1=-\n_0+\a+2\g.
\marke{reflect2}\ee

Let now the reflected ray (having the direction $\k_1$) be again
reflected by the mirror closing the optical system from the opposite
side. If this mirror is orthogonal to the optical axis, i.e. it has
normal $\n_0$, then the direction of the ray after this second
reflection is characterized by the vector
\be
\k'=\k_1-2(\k_1\n_0)\n_0=\n_0+\a+2\g.
\marke{secondRefl}\ee
Finally we see that the ray characterized by the vector angle $\a$
converts after double passage through the optical system to
the ray corresponding to the new vector angle $\a'$:
\be
\a\rightarrow\a'=\a+2\g.
\marke{twoRefl}\ee
Here the vector angle $\g$ characterizes inclination of the mirror
which reflected the ray. This mirror is only one lobe of the multilobe
mirror. In each double passage the ray is reflected from one of the
lobes chosen randomly.

\section{`Diffusion' of the inclination angle and the angular
distribution of rays.}\marke{SectMultiPass}

Assume that the multilobe mirror contains $s$ mirrors (lobes)
characterized by the vector angles $\{ \g_1, \g_2, \dots , \g_s\}$.
After each double passage the ray may be reflected by one of these
mirrors. If the mirror with number $i$, $i=1,2,\dots,s$ is chosen, the
direction of the ray is changed as follows:
\be
\a\rightarrow\a'=\a+2\g_i,
\marke{ith}\ee
In each double passage the number $i$ is chosen randomly according
some probability distribution. In the simplest case the probabilities
of all lobes are equal, $p_i=1/s$. We shall consider only this case.

After $N$ double passages an initial inclination angle changes as
\be
\a\rightarrow\a'=\a+\e
\marke{Npass}\ee
where
$$
\e=\ks_1+\ks_2+\dots+\ks_N
$$
is a random variable equal to the sum of $N$ random variables $\ks_i$.
All variables $\ks_i$ have equal distributions. Each of them takes the
values $\{ 2\g_1, 2\g_2, \dots, 2\g_s\}$ with corresponding
probabilities. If the probabilities of all values are equal to each
other, $p_i=1/s$, then the expectation values, variances and
covariations for the components of each of the (vector) random
variables $\ks_i$ are
\ban
M\xi_x=\frac{1}{s}\sum_i 2\gamma_{ix},
&& M\xi_y=\frac{1}{s}\sum_i 2\gamma_{iy}, \\
D\xi_x=\frac{1}{s}\sum_i (2\gamma_{ix}-M\xi_x)^2,
&& D\xi_y=\frac{1}{s}\sum_i (2\gamma_{iy}-M\xi_y)^2, \\
{\rm cov}(\xi_x,\xi_y)
&=& \frac{1}{s}\sum_i (2\gamma_{ix}-M\xi_x)(2\gamma_{iy}-M\xi_y).
\ean
We shall assume for the aim of simplicity that the set of vectors
$\{ \g_i \}$ is symmetrical so that
$$
M\xi_x=M\xi_y=0, \quad D\xi_x=D\xi_y=\sigma^2, \quad
{\rm cov}(\xi_x,\xi_y) = 0.
$$
Then the $x$- and $y$- components of the random variables $\xi_i$ have
the same characteristics and are independent from each other. For the
random variable $\e$ (which is the sum of $\ks_i$) we have
$$
M\e=0, \quad D\eta_x=D\eta_y=N\,\sigma^2 \quad
{\rm cov}(\eta_x,\eta_y) = 0.
$$

Change of the angle described as in \eq{Npass} is nothing else than
Brownian walk of a point in a plane. If $N$ is large, then, owing
to the central limiting theorem of the probability theory, the
distribution in various values ${\bf r}=(x,y)$ of the random
variable $\e$ may be approximated by the normal distribution:
\be
p_N(x,y)dxdy
=\frac{1}{2\pi N\sigma^2}
\exp\left( -\frac{x^2+y^2}{2N\sigma^2}\right)\, dxdy.
\marke{normDistr}\ee
The probability distribution for the components of the vector angle
$\a'$ (for the given initial angle $\a$) is found readily. If we
denote these components by the same letters $x$, $y$, then this
probability has the form
$$
P_N(x,y|\a)dxdy
=\frac{1}{2\pi N\sigma^2}
\exp\left( -\frac{(x-\al_x)^2+(y-\al_y)^2}{2N\sigma^2}\right)\, dxdy.
$$
If an arbitrary initial distribution of the angles $\a$ is given, then
the final distribution is
$$
P_N(x,y) = \int P_N(x,y|\a) \, P(\a) d\al_x d\al_y.
$$

If $N\sigma^2=t$ is considered as time, then the distribution function
$P(t,x,y)=P_N(x,y)$ satisfies the diffusion equation:
\be
\partderiv{P}{t}=\frac{1}{2}
\left( \partderiv{^2P}{x^2}+\partderiv{^2P}{y^2}\right).
\marke{eqDiffus}\ee
In the preceding consideration is has been assumed that the angle
$\a'$ may become, as a result of the random process, arbitrarily
large. In reality it is restricted by finiteness of the aperture. This
may be taken into account by the corresponding boundary conditions for
the equation \eq{eqDiffus}.

\section{Angular distribution for finite aperture}

When after a number of double passages the inclination angle becomes
sufficiently large, the ray will be lost owing to the finiteness
of aperture. In the language of Brownian walk this means that the
point disappears and takes no part in further random process. If we
have initially many Brownian particles, part of them is lost because
of finite aperture so that the final number is less. In terms of the
probability distribution, its norm decreases with time.

At first glance, this argument leads to null conditions at the
boundary of some finite region in the plane of Brownian walk. Adding
this boundary condition to the equation \eq{eqDiffus}, we could find
the distribution taking into account finiteness of the aperture. In
the case of axial symmetry of the optical system the permitted region
might be a disc of the radius restricting the maximum absolute value
of the angle depending on the aperture. One is attempted to impose
null conditions at the angle equal to $\Gamma=a/L$ where $a$ is the
radius of the mirror and $L$ distance between the opposite mirrors of
the optical system.

This is however not so simple. Precisely speaking, no null boundary
condition may be formulated for the angle $\a'$. Yet we shall show
that this may be made approximately. The null initial condition must
be imposed at some critical angle $\Gamma_c$ depending on $\Gamma$ and
the angle $\sigma$ characterizing deflection at each double passage.

The precise formulation of the problem is possible only in terms of
the joint distribution $P(\al, \r)$ in the angle $\a$ of inclination of
the ray and position $\r$ of the point of reflection this ray at the
mirror. It must be required that this distribution function be zero
when $\r$ takes values corresponding to the edge of the mirror. The
problem is that the equation for the function $P(\al, \r)$ is
complicated and cannot be dealt efficiently.

Instead of this correct consideration we shall restrict ourselves by
the function $P(\al)$ and the diffusion equation for it. As for the
boundary conditions, we shall find, up to the order of magnitude, the
angle $\Gamma_c$ such that when this angle of inclination is reached,
the ray is with great probability lost in some time after this.

To find the critical angle $\Gamma_c$, we shall argue in the following
way. Let the inclination angle $\Gamma_0$ is reached in the course of
Brownian walk. If this angle leaves unchanged during next passages
through the system, then after each double passage the point of the
reflection at the mirror is replaced by $2L\Gamma_0$. Then after
$N_0=\Gamma/\Gamma_0$ double passages the point of reflection is
replaced by the value $2LN_0\Gamma_0=2L\Gamma=2a$ so that the ray will
certainly escape from the optical system.

However after reaching the value $\Gamma_0$ the inclination angle will
not stay constant but change in the course of Brownian walk. The
preceding conclusion that the ray will be lost in $N_0$ double
passages will be valid if the change of the angle is much less than
$\Gamma_0$. Making use of typical Brownian replacement
$\sqrt{N_0}\sigma$, we have inequality $\sqrt{N_0}\sigma \ll \Gamma_0$
as the condition that the ray will be lost. Equivalently this
condition may be written as $\Gamma_0 \gg \Gamma_c$ where
$$
\Gamma_c = (\Gamma\sigma^2)^{1/3}
$$
Vice versa, if $\Gamma_0 \ll \Gamma_c$ then the random changes of the
angle $\Gamma_0$ in the course of Brownian walk will radically change
it before the ray will be lost. In this case the fact that the angle
reached the value $\Gamma_0$ has no special significance.

We see that the angle $\Gamma_c$ is critical in the sense that
reaching this angle in the course of Brownian walk leads with great
probability to loss of the ray. This means that the distribution
function $P(\al)$ must be small for $|\al|=\Gamma_c$. Approximately we
may take it to be zero for such arguments. This means imposing null
boundary conditions at the boundary of the disc of radius $\Gamma_c$.

Thus, the diffusion equation \eq{eqDiffus} should be solved with the
null boundary conditions at the boundary of the disc of radius
$\Gamma_c$:
$$
P(t,x,y)=0 \;\; \mbox{ when } \;\; x^2+y^2=\Gamma_c^2.
$$
This solution is easily obtained with the help of the separation
of variables in the polar coordinates. It has the form
\ba
P(t,r,\varphi)&=&P(t,x,y)\nonumber\\
&=&\sum_{n=0}\sum_{m=1}
e^{-\omega_{mn}^2\,t}\, J_n(k_{mn}r) \, \al_{mn} \cos n\varphi
\nonumber\\
&+&\sum_{n=1}\sum_{m=1}
e^{-\omega_{mn}^2\,t}\, J_n(k_{mn}r) \, \beta_{mn} \sin n\varphi
\marke{solut}\ea
where the following notations are used:
$$
\omega_{mn}^2=\frac 12 \, k_{mn}^2, \quad
k_{mn}=\mu_{mn}/\Gamma_c, \quad
\mu_{mn} \mbox{ are the roots of the Bessel function } J_n(r).
$$
The coefficients $\al_{mn}$, $\beta_{mn}$ are determined by
initial conditions as follows:
$$
P(0,r,\varphi)
=\sum_{n=0}\sum_{m=1}
(\al_{mn} \cos n\varphi+\beta_{mn} \sin n\varphi) J_n(k_{mn}r).
$$

Asymptotically (at large $t=N\sigma^2$) the first term in the
formula \eq{solut} dominates because it corresponds to the
minimal root of the Bessel functions and therefore minimal
exponent. Therefore after a large number of double passages
\be
p(t,r,\varphi)=P(t,x,y)=P_N(x,y)
\approx  \al_{10}e^{-\omega_{10}^2\,t}\, J_0(\mu_{10}r/\Gamma_c) .
\marke{asympSolut}\ee
It is only numerical factor $\al_{10}$ which depends on initial
conditions in this expression. Therefore, asymptotic (for large
$N$) angular distribution of the radiation issued by the optical
system with the multilobe mirror does not depend on initial
conditions and is described by the Bessel function of zero
order.\footnote{In the very special case when initial conditions
correspond precisely zero coefficient $\al_{10}$, the asymptotic
distribution is determined by the next term of the sum
corresponding to the larger root of the Bessel function.}

\section{Conclusion}

In the present paper we have considered (in the framework of the
geometric optics) the rays of light in an optical system having an
usual plane mirror from one side and a multilobe mirror from another
side. We evaluated the probability that the outcoming ray has a
definite inclination angle. It was shown that the reflection of a ray
from one of many lobes of the multilobe mirror inclines it in one of
many directions. As a result, the inclination angle undergoes
Brownian walk. The resulting probability distribution satisfies then
the diffusion equation, with the number of passages playing the role
of time.

If there are many input rays, then the same function which describes
the probability distribution gives the angular distribution of the
outcoming rays. It was shown that the distribution alters when the
number of passages of the ray through the optical system increases.
Asymptotically for very large number of passages the distribution
tends to some universal distribution which possesses the same symmetry
as the lobes of the multilobe mirror (axial symmetry in the considered
case).

If a (modeless) laser is constructed on the basis of such an optical
system, its radiation will have distribution identical with the
asymptotic distribution we have found. The angular distribution of 
the multilobe mirror laser proves to be stable under perturbations, 
therefore the distribution \eq{asympSolut} maintains in the non-ideal 
conditions (for example with inhomogeneous active medium). This 
explains why the multilobe mirror laser have smooth
(homogeneous) angular distribution \cite{MultilobeMir,Exper}.

Several methods could be applied to increase the homogeneity of active
multilobe mirror lasers. Among them 1) the increasing of the Fresnel
number (i. e. the increasing of the laser aperture and the decreasing
of the  cavity  length and wavelength), 2) the  increasing of the
number of lobes (planes) of the multilobe mirror, 3) the application
of the active medium with sufficient broad bandwidth (for example
Neodymium glass active elements). The above developed method of
calculations may be generalized to include these modifications
increasing the homogeneity of a laser beam.



\end{document}